\documentclass[a4paper]{article}

\usepackage{INTERSPEECH2018, multirow}

\title{Deep Time Delay Neural Network for Speech Enhancement with Full Data Learning}
\name{Cunhang Fan$^{1,3}$, Bin Liu$^1$, Jianhua Tao$^{1,2,3}$, Jiangyan Yi$^{1}$, Zhengqi Wen$^1$, Leichao Song$^{1,3}$}
\address{ $^1$National Laboratory of Pattern Recognition, Institute of Automation, Chinese Academy of Sciences, Beijing, China\\
	$^2$CAS Center for Excellence in Brain Science and Intelligence Technology, Beijing, China\\
	$^3$School of Artificial Intelligence, University of Chinese Academy of Sciences, Beijing, China}
\email{\{cunhang.fan, liubin, jhtao, yibin.zheng, jiangyan.yi, zqwen\}\@nlpr.ia.ac.cn}

\begin{document}

\maketitle
\begin{abstract}
  Recurrent neural networks (RNNs) have shown significant improvements in recent years for speech enhancement. However, the model complexity and inference time cost of RNNs are much higher than deep feed-forward neural networks (DNNs). Therefore, these limit the applications of speech enhancement. This paper proposes a deep time delay neural network (TDNN) for speech enhancement with full data learning. The TDNN has excellent potential for capturing long range temporal contexts, which utilizes a modular and incremental design. Besides, the TDNN preserves the feed-forward structure so that its inference cost is comparable to standard DNN. To make full use of the training data, we propose a full data learning method for speech enhancement. More specifically, we not only use the noisy-to-clean (input-to-target) to train the enhanced model, but also the clean-to-clean and noise-to-silence data. Therefore, all of the training data can be used to train the enhanced model. Our experiments are conducted on TIMIT dataset. Experimental results show that our proposed method could achieve a better performance than DNN and comparable even better performance than BLSTM. Meanwhile, compared with the BLSTM, the proposed method drastically reduce the inference time.
\end{abstract}

\section{Introduction}

The performance of speech processing applications in domains such as automatic speech and speaker recognition, speech coding and hearing aids degrades significantly when the test data is noisy \cite{Li2014An}. Therefore, speech enhancement is extremely essential for speech processing applications. 


Over the years, motivated by the success of deep learning, researchers have developed many deep learning techniques for speech enhancement, such as deep feed-forward neural networks (DNNs) \cite{xu2015regression,Wang2013Towards}, deep denoising auto-encoders \cite{lu2013speech}, recurrent neural networks (RNNs) \cite{weninger2015speech,Subramanian2018} and convolutional neural networks (CNNs) \cite{Park2017A}. RNNs \cite{robinson1988static} can make full use of historical information and capture long-term dependencies in sequential data by means of a dynamically changing context window over all of the sequence, especially for long short-term memory (LSTM) \cite{hochreiter1997long} and bidirectional LSTM (BLSTM) \cite{schuster1997bidirectional}. However, because of the recurrent connections in the RNNs, compared with DNNs, the model complexity of RNNs is much higher. In addition, the training and inference time cost are exceedingly longer than DNNs.

Different from DNNs that only use a fixed and small context window as the input, the time delay neural network (TDNN) has an effective ability in capturing long range temporal contexts \cite{waibel1989phoneme,peddinti2015time,Zheng2018}. The architecture of TDNN applies a modular and incremental design in order to create a larger network from sub-components \cite{waibel1989modular}. What's more, the training and inference time of TDNN are comparable with DNNs because the TDNN preserves the feed-forward structure \cite{peddinti2015time}. 

Usually, speech enhancement is a part of the front-end for speech processing applications. Therefore, the model of speech enhancement should be low latency. However, it is very difficult to reduce the model complexity while ensuring performance. In order to address this issue, in this paper, we propose a full data learning method for speech enhancement based on the TDNN. To improve the performance of speech enhancement, we investigate the influences of different temporal context windows in each TDNN layer. Most of the speech enhancement methods only use the noisy-to-clean (input-to-target) data to train their model but the clean-to-clean data and noise-to-silence data are not used. In order to make full use of the training data, the full data learning is proposed. More specifically, we not only use the noisy data, but also the clean and noise data. The target of speech enhancement is to remove noise from noisy speech. If the input of the system is clean data, the output should be same as the input. Conversely, if the input of the system is noise data, the output should be the silence. Therefore, we use this trick to improve the performance of speech enhancement. To our best knowledge, it is the first time to use this trick for speech enhancement. 

The rest of this paper is organized as follows. In section 2, our baseline DNN/BLSTM based speech enhancement system is introduced. Section 3 describes the proposed deep TDNN speech enhancement system. Section 4 shows detailed experiments and results. Section 5 draws conclusions.

\begin{figure*}[t]
	\centering
	\begin{minipage}[t]{0.78\textwidth}
		\includegraphics[width=\linewidth]{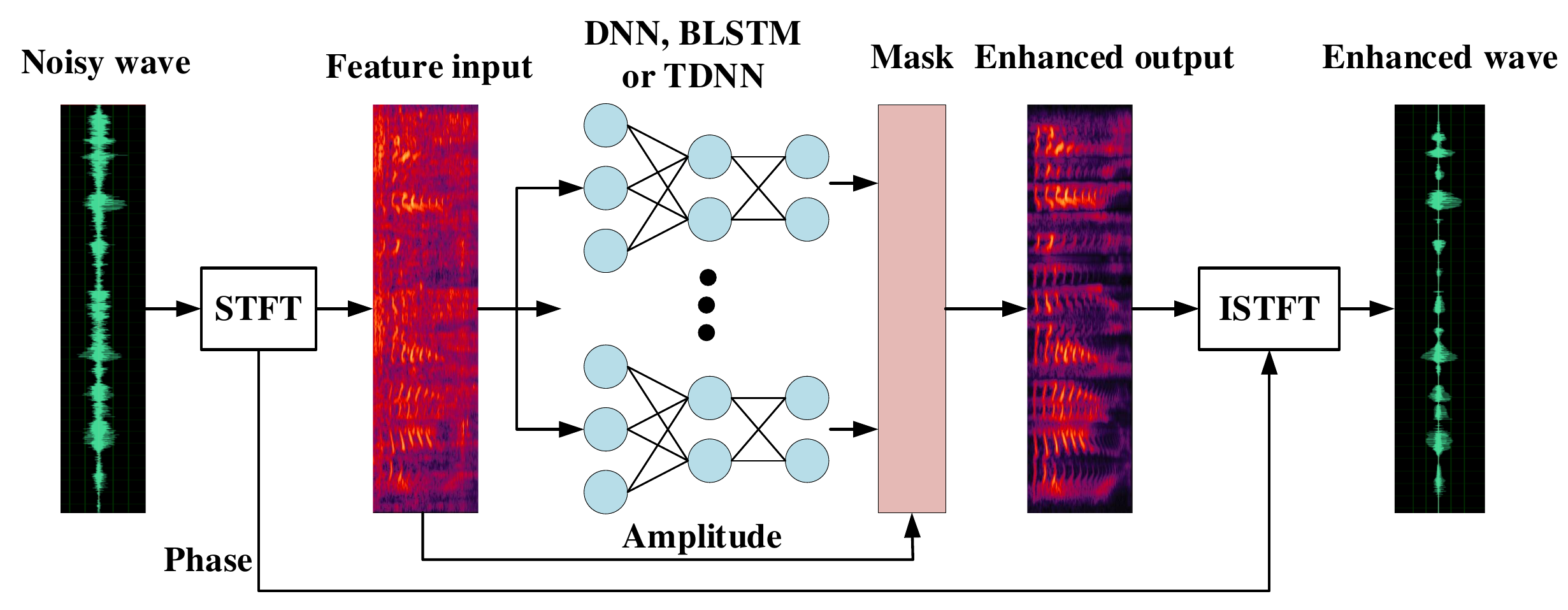}
	\end{minipage}
	\caption{The architecture of DNN, BLSTM or TDNN based speech enhancement system.}
	\label{fig:DNN-architecture}
\end{figure*}

\section{DNN/BLSTM based speech enhancement system}


The aim of speech enhancement is to remove the additive noise signal from the noisy speech. The noisy speech can be represented as:
\begin{equation}
y(t) = x(t)+n(t)
\label{eq1}
\end{equation}
where \(y(t)\), \(x(t)\) and \(n(t)\) note the noisy, clean and noise signals, respectively. The corresponding short-time Fourier transformation (STFT) of those signals are \(Y(t,f)\), \(X(t,f)\) and \(N(t,f)\). The following relationship is still satisfied after STFT
\begin{equation}
Y(t,f) = X(t,f)+N(t,f)
\label{eq22}
\end{equation}
As for speech separation task, it is well known that mask based speech separation method can obtain a better result \cite{fan2019discriminitive,fan2020end,fan2020gated,fan2020joint,wang2018supervised}. Similarly, we apply the mask \(M(t,f)\) for speech enhancement.


According to the commonly used masking method, the estimated magnitude  \(|\widetilde{X}(t,f)|\)  of clean speech can be estimated by
\begin{equation}
|\widetilde{X}(t,f)|=|Y(t,f) |\odot{M(t,f)}
\label{eq3}
\end{equation}
where \(\odot\) indicates element-wise multiplication, \(M(t,f)\) is the mask estimated by DNN or BLSTM. In this study, both DNN and BLSTM are employed as the baseline systems. Finally, the estimated magnitude  \(|\widetilde{X}(t,f)|\) and the phase of noisy speech is used to reconstruct enhanced signal by inverse STFT (ISTFT). 

Fig.\ref{fig:DNN-architecture} shows the overview of the streaming speech enhancement architecture using DNN or BLSTM. The noisy wave is first transformed into time-frequency domain by short-time Fourier transformation (STFT). Then the amplitude spectrum is used as the input feature of DNN or BLSTM. A DNN or BLSTM is applied to model the mapping between the input amplitude features and the output target masks. Next, the target output can be acquired by masks and the input. Finally, the enhanced output and the phase of noisy signal are used to reconstruct the signal by inverse STFT (ISTFT). 

\section{The proposed speech enhancement system}

The proposed deep TDNN based speech enhancement system is similar to DNN or BLSTM that is described in Section2, except we replace the DNN/BLSTM with deep TDNN. Besides, in order to make full use of training data, we propose the full data learning method, which applies the clean and noise data to fine-tuning the enhanced model. In this section, we will firstly review the basic architecture of TDNN, and then demonstrate how to make full use of the clean and noise data to fine-tune the model. Finally, the training procedure of the proposed model is introduced.

\subsection{Deep TDNN architecture}

\begin{figure}[t]
	\centering
	\begin{minipage}[t]{0.35\textwidth}
		\includegraphics[width=\linewidth]{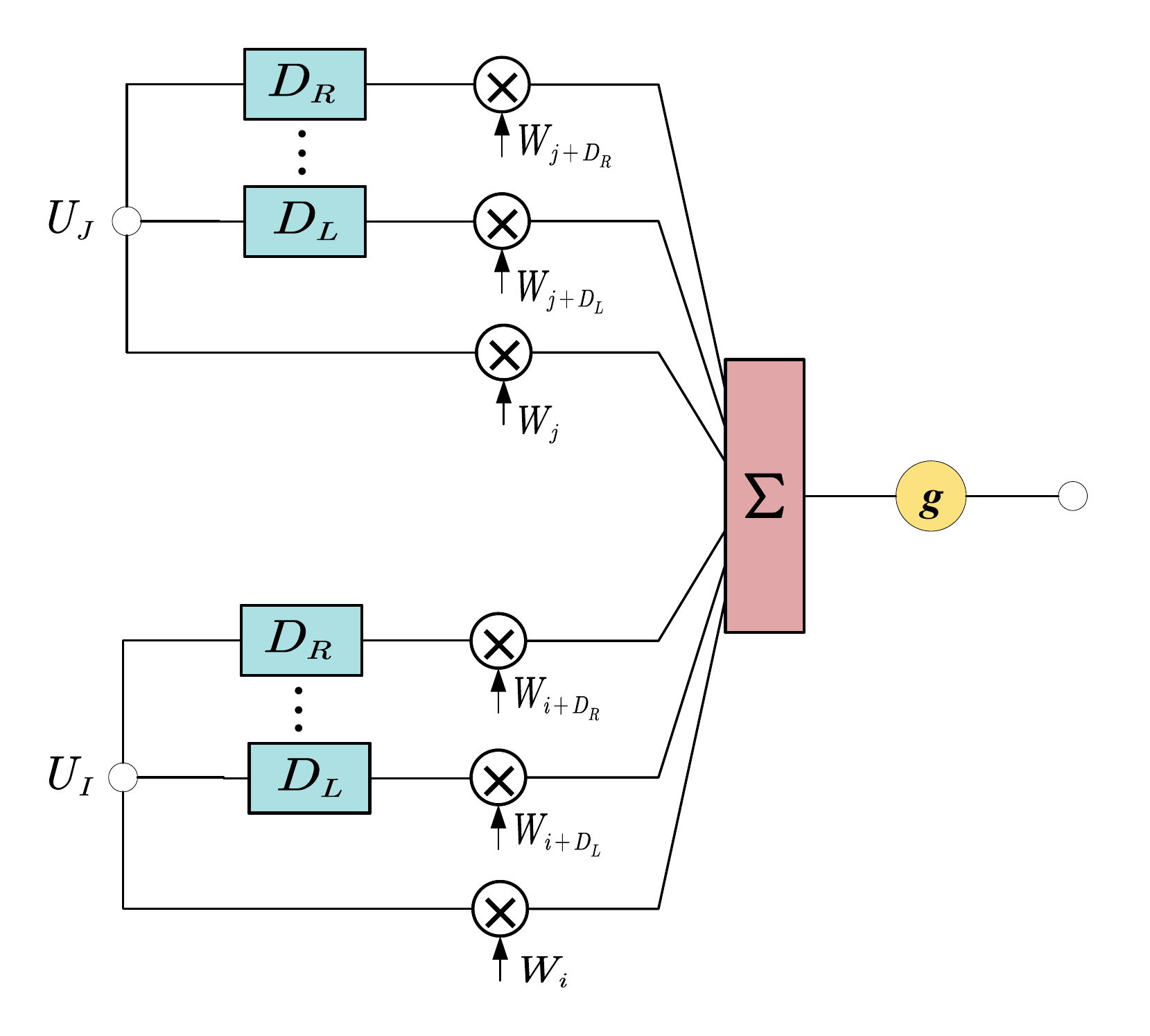}
	\end{minipage}
	\caption{A time delay neural network (TDNN) unit.}
	\label{fig:TDNN-architecture}
\end{figure}

As for the basic unit in DNN, it computes the weighted sum of its inputs and then passes them through a nonlinear function, such as sigmoid or tanh function. However, in the TDNN \cite{waibel1989modular}, this basic unit is modified by introducing delays \(D_L\) through \(D_R\) as shown in Figure \ref{fig:TDNN-architecture}. The \(j\)-th inputs of such a unit now will be multiplied by several weights. In this way, a TDNN unit has the ability to relate and compare current input to the past history of events. Therefore, the TDNN can acquire more effective information from the noisy speech so that it can help improve the performance of speech enhancement. The activation function for node \(i\) at time \(t\) in such a network is given by:
\begin{equation}
o_i^t = g(\sum_{j=1}^{i-1}\sum_{k=D_L}^{D_R}o_j^{t-k}w_{ijk})
\label{eq4}
\end{equation}
where \(o_i^t\) is the output of node \(i\) at time \(t\), \(w_{ijk}\) is the connection strength to node \(i\) from output of node \(j\) at time \(t-k\), and \(g(*)\) is the activation function. 


\subsection{Full data learning method and loss function}

\begin{figure}[ht]
	\centering
	\begin{minipage}[t]{0.35\textwidth}
		\includegraphics[width=\linewidth]{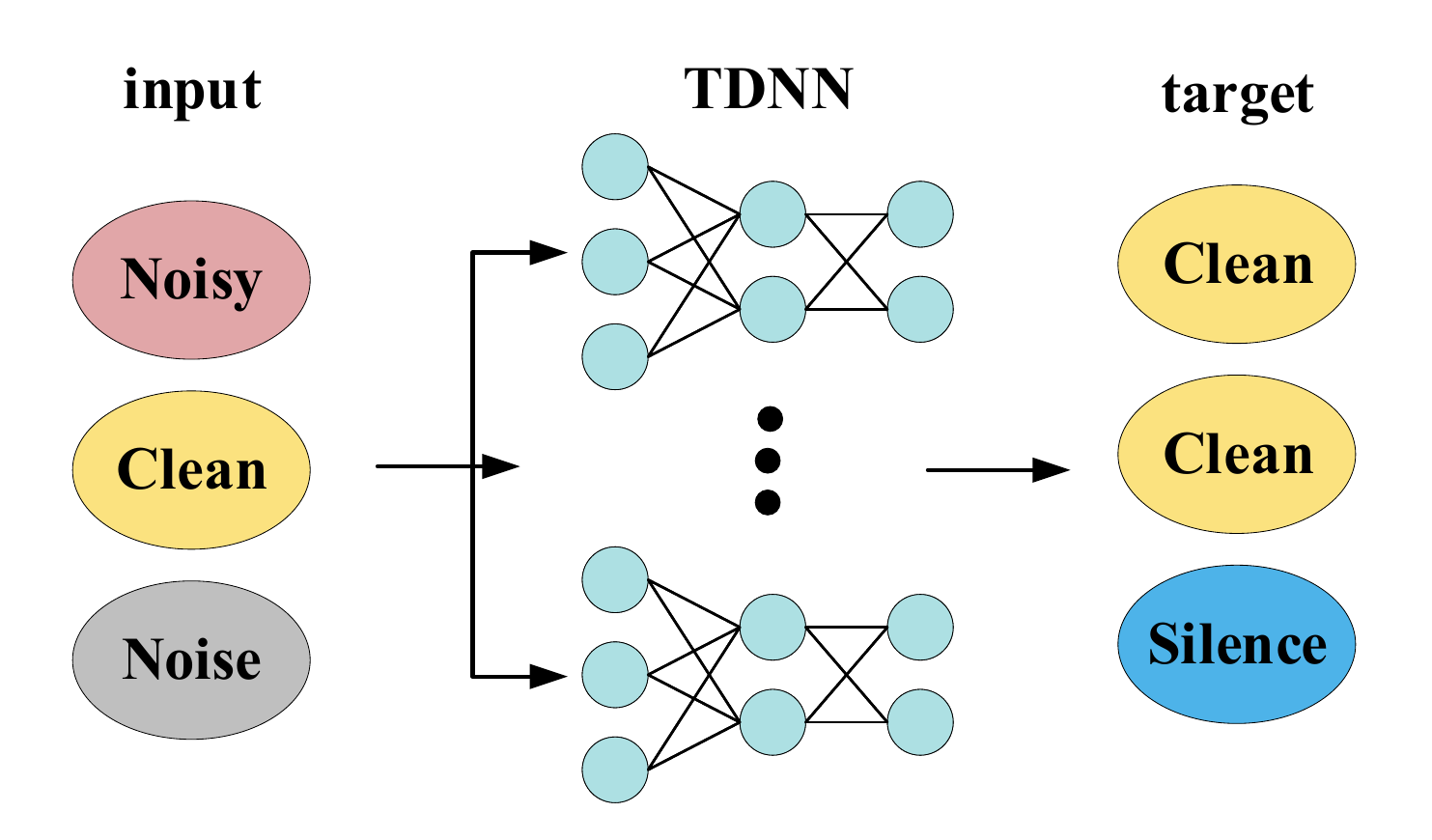}
	\end{minipage}
	\caption{The overview of our proposed full data learning. Using noisy-to-clean, clean-to-clean and noise-to-silence (input-to-target) to train the speech enhancement system.}
	\label{fig:clean-noise}
\end{figure}

Most of the speech enhancement systems only use the noisy features as the input, and clean features are as the output. However, the clean-clean and noise-silence are not used, which does not make full use of the training data. In order to address this problem and improve the performance of speech enhancement, we propose a full data learning method. It not only uses the noisy features as the input, but also clean and noise features. The target of speech enhancement is to remove the noise from an observed signal recorded in noisy environment. However, if the input of the enhanced model is clean signals, the output should be same as the input. Conversely, if noise is the input of the enhanced model, the output should be silence. In other words, we make full use of the noisy-to-clean, clean-to-clean and noise-to-silence (input-to-target) to train the enhanced model as shown in Fig. \ref{fig:clean-noise}.


The ideal amplitude mask (IAM) \(M{_{iam}}(t,f)\) \cite{wang2018supervised} which is a widely-accepted mask for speech enhancement is utilized in this paper.
\begin{equation}
M_{iam}(t,f)=\frac{|X(t,f)|}{|Y(t,f)|}
\label{eq6}
\end{equation}
The loss function is the mean square error (MSE) between the estimated magnitude and true magnitude:
\begin{equation}
J=\frac{1}{TF}\sum|||Y(t,f)|\odot{\widetilde{M}_{iam}(t,f)}-|X(t,f)|||_F^2
\label{eq7}
\end{equation}
where \(\widetilde{M}_{iam}(t,f)\) is the estimated mask, \(\odot\) indicates element-wise multiplication.

The training procedure for our proposed speech enhancement system is as follows:
\begin{itemize}
	\item Use the noisy-clean training data to train a deep TDNN based speech enhancement model.
	\item Fine tuning the enhanced model with clean-clean training data.
	\item Fine tuning the enhanced model with noise-silence training datas.
	\item Use the noisy-clean training data to optimize the model, again.
\end{itemize}

%

\section{Experiments and results}

\subsection{Dataset}

The experiments are conducted using the TIMIT database \cite{garofolo1993darpa}. We create the training, validation and test sets in the same manner. The noises use in the training and validation sets include 100 different noise types, which can be download from the website\footnote{http://web.cse.ohio-state.edu/pnl/corpus/HuNonspeech/HuCorpus.html}. The training and validation sets are generated by randomly selecting speakers and utterances from the TIMIT training set. They are mixed with this noise database at 6 signal-to-noise ratio (SNR) (-5, 0, 5, 10, 15 and 20 dB). The training and validation sets have about 21726 and 6006 utterances, respectively. As for the test set, besides the 100 different noise types, twelve unseen noises are used, which are from NISEX-92 dataset \cite{varga1993assessment}. Same to the training set, they are mixed with the utterances from TIMIT test set at 6 SNR (-5, 0, 5, 10, 15 and 20 dB). The test set contains 10086 utterances. We use the validation set to select the best model from all epochs.

\subsection{Experimental setups}

The sampling rate of all generated data is 8 kHz. The 129-dim spectral magnitudes of the noisy speech are used as the input features, which are computed using a STFT with 32 ms length hamming window and 16 ms window shift. ReLU function is employed as the activation function for deep TDNN training. All the TDNN based systems have four hidden layers with 256 nodes each layer in this work. When we only use the noisy-to-clean data to train the model, there is 30 epochs. The clean-to-clean and noise-to-silence are all used to fine tuning the enhanced model with 5 epochs.


In this paper, the learning rate of all models are initialized as 0.0005 and scaled down by 0.7 when the training objective function value increased on the development set. Our models are optimized with the Adam algorithm. All models are trained by one Tesla K80.

\subsection{Baseline models}

For comparison purposes, two types of systems (described in Section 2), which are DNN and BLSTM respectively, are built. These two systems serve as two baselines in this paper.

\begin{itemize}
	\item DNN: As for the DNN based system, there are four hidden layers and each layer has 256 nodes, which is same as the TDNN method.
	\item BLSTM: For BLSTM based system, we use three BLSTM layers with 256 memory blocks for each layer. It contains random dropouts with a dropout rate 0.5.
\end{itemize}

\subsection{Evaluation metrics}

In order to evaluate the quality of the enhanced speech, we compute the following objective measures: the perceptual evaluation of speech quality (PESQ) \cite{rix2002perceptual} measure, the short-time objective intelligibility (STOI) measure \cite{taal2010short} and the signal-to-distortion ratio (SDR) measure \cite{vincent2006performance}.

\subsection{Experimental results}

\setlength{\tabcolsep}{1.5mm}{
	\begin{table}[]
		\caption{The configuration of layerwise temporal context windows for different speech enhancement methods.}
		\label{tab:configure}
		\centering
		\begin{tabular}{cc|ccccc}
			\hline
			\multirow{2}{*}{Methods}&
			\multirow{1}{*}{Network}&
			\multicolumn{5}{c}{Layerwise Context}\\
			\cline{3-7}
			&Context & 1 & 2 & 3 & 4 & 5 \\
			\hline
			DNN & [-8,8]& [-8,8]& 0 & 0& 0 & 0 \\
			\hline
			TDNN-A & [-11,11]& [-4,4]& [-3,3] & [-2,2]& [-2,2] & 0 \\
			TDNN-B & [-10,10]& [-2,2]& [-2,2] & [-2,2]& [-4,4] & 0 \\
			TDNN-C & [-9,9]& [-2,2]& [-1,1] & [-2,2]& [-4,4] & 0 \\
			TDNN-D & [-8,8]& [-2,2]& [-2,2] & [-2,2]& [-2,2] & 0 \\
			TDNN-E & [-7,7]& [-1,1]& [-2,2] & [-2,2]& [-2,2] & 0 \\
			TDNN-F & [-6,6]& [-1,1]& [-1,1] & [-2,2]& [-2,2] & 0 \\
			\hline
		\end{tabular}
\end{table}}

\setlength{\tabcolsep}{1.2mm}{
	\begin{table*}[t]
		\caption{The performance of different systems on unseen, seen and average (AVG.) condition. (+clean+noise) means the proposed full data learning method that uses the clean-to-clean and noise-to-silence to fine tuning the enhanced model.}
		\label{tab:results1}
		\centering
		\begin{tabular}{c|ccc|ccc|ccc|c}
			\hline
			\multirow{2}{*}{Methods}                                  
			& \multicolumn{3}{c|}{unseen}                    & \multicolumn{3}{c|}{seen} & \multicolumn{3}{c|}{AVG.} &Inference  \\ \cline{2-10} 
			& PESQ          & STOI(\%)       & SDR           & PESQ          & STOI(\%)       & SDR  & PESQ          & STOI(\%)       & SDR      &{time(ms)}  \\ \hline
			noisy                    & 2.46          & 75.88          & 7.7           & 2.37          & 83.16          & 7.7  &2.38 &82.40 & 7.7    &-      \\ \hline
			DNN(baseline)            & 2.76          & 78.77          & 13.7          & 2.84          & 86.66          & 14.4  &2.83 &85.83 & 14.3  &13.7     \\ 
			BLSTM(baseline)          & {2.90}  & {79.29} & {15.4} & {2.99} & {87.67} & {15.8} &{2.98} &{86.79} &{15.8} &429.8 \\ \hline
			TDNN-A(proposed)         & 2.81          & 78.85          & 14.1          & 2.88          & 86.89          & 14.7   &2.87 &86.05 & 14.6   &14.5    \\ 
			TDNN-B(proposed)         & 2.81          & 78.98          & \textbf{14.3} & 2.88          & 86.95          & {14.9} &2.87 &86.12 &14.8  &{14.6} \\ 
			TDNN-C(proposed)         & 2.80          & 79.14          & 14.1          & 2.88          & 86.98          & {14.9} &2.87 &86.16 &14.8 &15.3 \\ 
			TDNN-D(proposed)         & 2.82          & 79.55          & 14.2          & \textbf{2.92} & 87.34          & {14.9} &\textbf{2.91} &86.52 &14.8 &14.2 \\ 
			TDNN-E(proposed)         & \textbf{2.83} & 79.42          & 14.2          & 2.91          & 87.31          & {14.9} &2.90 &86.48 &14.8 &12.9 \\ 
			TDNN-F(proposed)         & 2.82          & {79.76} & 14.1          & \textbf{2.92} & {87.42} & {14.9} &\textbf{2.91} &86.62 & 14.8 &14.5 \\ \hline
			{TDNN-F}         & \multirow{2}{*}{2.81}          & \multirow{2}{*}{\textbf{79.78}} & \multirow{2}{*}{14.2}          & \multirow{2}{*}{\textbf{2.92}}          & \multirow{2}{*}{\textbf{87.61}} & \multirow{2}{*}{\textbf{15.0}} &\multirow{2}{*}{\textbf{2.91}} &\multirow{2}{*}{\textbf{86.79}} &\multirow{2}{*}{\textbf{14.9}} &\multirow{2}{*}{14.5} \\ 
			(+clean+noise)(proposed) &&&&&&&&&& \\ \hline
		\end{tabular}
\end{table*}}

\setlength{\tabcolsep}{1.0mm}{
	\begin{table*}[]
		\caption{The performance of different systems for different SNRs.}
		\label{tab:results2}
		\centering
		\begin{tabular}{c|cccccc|cccccc|cccccc}
			\hline
			& \multicolumn{6}{c|}{PESQ}                                                                     & \multicolumn{6}{c|}{STOI(\%)}                                                                       & \multicolumn{6}{c}{SDR}                                                                     \\ \hline
			SNR(dB) & -5            & 0             & 5             & 10            & 15            & 20            & -5             & 0              & 5              & 10             & 15             & 20             & -5           & 0             & 5             & 10            & 15            & 20            \\ \hline
			DNN     & 2.21          & 2.50          & 2.75          & 2.96          & 3.17          & 3.41          & 74.18          & 80.43          & 85.22          & 89.05          & 92.01          & 94.12          & 7.2          & 10.1          & 12.8          & 15.6          & 18.6          & 21.7          \\ 
			BLSTM   & {2.26} & {2.61} & {2.90} & {3.16} & {3.37} & {3.56} & {76.03} & {81.60} & {86.04} & {89.69} & {92.58} & {94.82} & {8.4} & {11.2} & {14.0} & {17.0} & {20.2} & {23.6} \\ \hline
			TDNN-A  & 2.21          & 2.53          & 2.79          & 3.02          & 3.23          & 3.44          & 74.29          & 80.63          & 85.45          & 89.26          & 92.22          & 94.43          & 7.2          & 10.2          & 13.1          & 16.0          & 19.1          & 22.3          \\ 
			TDNN-B  & 2.18          & 2.51          & 2.79          & 3.03          & 3.25          & 3.47          & 74.54          & 80.80          & 85.54          & 89.25          & 92.19          & 94.37          & 7.4          & {10.4} & 13.2          & 16.1          & \textbf{19.3} & \textbf{22.5} \\ 
			TDNN-C  & 2.19          & 2.51          & 2.79          & 3.03          & 3.25          & 3.47          & 74.62          & 80.87          & 85.60          & 89.31          & 92.20          & 94.37          & 7.4          & 10.3          & 13.2          & 16.1          & 19.2          & \textbf{22.5} \\ 
			TDNN-D  & \textbf{2.24} & \textbf{2.56} & \textbf{2.83} & {3.06} & \textbf{3.28} & 3.48          & 75.12          & 81.33          & 86.01          & 89.64          & 92.45          & 94.56          & 7.4          & {10.4} & \textbf{13.3} & {16.2} & \textbf{19.3} & \textbf{22.5} \\ 
			TDNN-E  & 2.22          & 2.55          & 2.82          & 3.05          & 3.27          & 3.48          & 75.11          & 81.29          & 85.95          & 89.61          & 92.43          & 94.51          & \textbf{7.5} & {10.4} & 13.2          & {16.2} & \textbf{19.3} & \textbf{22.5} \\
			TDNN-F  & 2.22          & 2.55          & \textbf{2.83} & {3.06} & \textbf{3.28} & \textbf{3.50} & {75.23} & {81.44} & {86.12} & {89.76} & {92.55} & {94.62} & 7.4          & {10.4} & \textbf{13.3} & {16.2} & \textbf{19.3} & \textbf{22.5} \\ \hline
			{TDNN-F}  & 	\multirow{2}{*}{2.22}          & 	\multirow{2}{*}{2.55}          & 	\multirow{2}{*}{\textbf{2.83}} & 	\multirow{2}{*}{\textbf{3.07}} & 	\multirow{2}{*}{\textbf{3.28}} & 	\multirow{2}{*}{\textbf{3.50}} & 	\multirow{2}{*}{\textbf{75.44}} & 	\multirow{2}{*}{\textbf{81.64}} & 	\multirow{2}{*}{\textbf{86.29}} & 	\multirow{2}{*}{\textbf{89.89}} & 	\multirow{2}{*}{\textbf{92.69}} & 	\multirow{2}{*}{\textbf{94.79}} & 	\multirow{2}{*}{\textbf{7.5}}          & 	\multirow{2}{*}{\textbf{10.5}} & 	\multirow{2}{*}{\textbf{13.3}} & 	\multirow{2}{*}{\textbf{16.3}} & 	\multirow{2}{*}{\textbf{19.3}} & 	\multirow{2}{*}{\textbf{22.5}} \\ 
			(+noise+clean) &&&&&&&&&&&&&&&&&& \\\hline
		\end{tabular}
	\end{table*}
}


%
%

Table~\ref{tab:results1} shows the performance of different systems with various layerwise temporal context windows on unseen, seen and average (AVG.) condition. To our best knowledge, there is no TDNN-based speech enhancement system. Therefore, we start exploration from a deep TDNN system. Table~\ref{tab:results2} shows the performance of different systems for different SNRs. Table~\ref{tab:configure} shows the configuration of layerwise temporal context windows for different speech enhancement methods. (+clean+noise) means the proposed full data learning method that uses the clean-to-clean and noise-to-silence to fine tuning the enhanced model.

\subsubsection{The effectiveness of the proposed TDNN method}

From Table~\ref{tab:results1}, we can find that no matter unseen or seen case, our proposed TDNN-based methods all achieve better performance than the DNN-based method in PESQ, STOI and SDR measures. Besides, system TDNN-D, TDNN-E and TDNN-F achieves comparable performance to BLSTM, even better than BLSTM (STOI measure for unseen case). In addition, we also show the inference time of different speech enhancement methods in the last column in Table~\ref{tab:results1}. The proposed TDNN-based methods use extremely less inference time than BLSTM and is comparable to DNN. These results suggest that our proposed TDNN-based speech enhancement models are very fast and the proposed method can reduce the model complexity while ensuring the performance.



\subsubsection{Evaluation of TDNN with full data learning method}

In order to make full use of the training data, the full data learning is proposed, which not only uses the noisy features as the input, but also clean and noise features. The results of full data learning are shown in the last row of Table~\ref{tab:results1} and Table~\ref{tab:results2}. From these results, we can know that when the enhanced model is optimized by clean-to-clean and noise-to-silence data, the performance of speech enhancement can be improved in most of cases, especially for the STOI measure. More specifically, besides the PESQ measure in unseen condition, the other results are all improved. In addition, form Table~\ref{tab:results2}, we can find that the proposed fine tuning method can improve the enhancement performance in objective measures no matter what SNRs. These results indicate the effectiveness of the proposed full data learning method. Therefore, using clean-to-clean and noise-to-silence to fine tuning the enhanced model can improve the performance of speech enhancement.



\section{Conclusions}

In this work, we propose a deep TDNN based method for speech enhancement, which requires low distortion and latency. The TDNN has an excellent potential for capturing long range temporal contexts, and its inference cost comparable to standard DNN. In order to make full use of training data, the full data learning is proposed, which uses clean-to-clean and noise-to-silence to fine tuning the enhanced model. It means that if the input of the enhanced model is clean signals, the output should be same as the input. Conversely, if noise is the input of the enhanced model, the output should be silence. Our experimental results show that the proposed speech enhancement method could achieve a better performance than DNN and comparable even better performance to BLSTM. In addition, compared with the BLSTM, the proposed method drastically reduce the enhanced generation time. 

%

\section{Acknowledgements}

This work is supported by the National Key Research \& Development Plan of China (No.2017YFB1002802), the National Natural Science Foundation of China (NSFC) (No.61831022, No.61901473, No.61771472, No.61773379 ) and Inria-CAS Joint Research Project (No.173211KYSB20190049).

\vfill\pagebreak

\bibliographystyle{IEEEtran}

\bibliography{mybib}


\end{document}